# NiTi Single Crystal Growth by Micro-Pulling-Down Method: Experimental Setup and Material Characterization

T. Sieweke[1,2], C. Luther[3], M. Wortmann[3,4], L. Schnatmann[3], F. Werner[5], O. Kuschel[4], L. Bondzio[3], I. Ennen[3], J. Bünte[3], K. Rott[3], O. Oluwabi[5], M. Loewenich[6], J. Wollschläger[4], A. Hütten[3], J. Frenzel[5], G. Schierning[1,2,7] and A. Kunzmann[1,2,]*

1. Institute for Energy and Materials Processes - Applied Quantum Materials, University of Duisburg-Essen, 47057 Duisburg, Germany
2. Research Center Future Energy Materials and Systems, Research Alliance Ruhr, 44780 Bochum, Germany
3. Thin Films & Physics of Nanostructures, Bielefeld University, 33615 Bielefeld, Germany
4. Institute of Physics, University Osnabrück, 49076 Osnabrück, Germany
5. Institute for Materials, Ruhr University Bochum, 44780 Bochum, Germany
6. Institute for Energy and Materials Processes - Reactive Fluids, University of Duisburg-Essen, 47057 Duisburg, Germany
7. Center for Nanointegration Duisburg-Essen (CENIDE) and Nano Energie Technik Zentrum (NETZ), 47057 Duisburg, Germany
* alexander.kunzmann@uni-due.de

## Abstract

Nickel-titanium that has an austenite to martensite phase transition has been studied extensively in the past as a shape memory alloy, but a lot remains to be learned from such phase transitions [1]. However, single crystals are needed for a detailed characterization of the emerging phase transition. In order to produce NiTi single crystals for research purposes, we have set up a micro-pulling-down (µPD) apparatus. The µPD process is a fast and flexible method for the fabrication of small single crystals [2]. The apparatus is operated in vacuum. By pulling the crystal down through a hole in the crucible bottom, it is possible to reduce oxygen contamination, since oxides float on top of the melt due to their low density. Here we present a detailed characterization of as-grown NiTi crystals by electron backscatter diffraction (EBSD), scanning electron microscopy (SEM), energy dispersive X-ray spectroscopy (EDX), X-ray photoelectron spectroscopy (XPS), hot gas extraction method and differential scanning calorimetry (DSC). The characteristics of the phase transition in NiTi are very sensitive to dopants and alloying. The µPD method facilitates the introduction of different doping elements into the crystal.

## 1. Introduction

NiTi, also known as Nitinol, goes back to its discovery at the US Naval Ordnance Laboratory [3] and is a very well-researched shape memory alloy with numerous possible applications, particularly in medical technology [4] and actuation [5], as well as in the aerospace industry [6]. The thermal shape recovery during heating enables the application in actuators. Its unique pseudoelasticity associated with the martensitic phase transition of NiTi is exploited for medical applications such as stents and it exhibits a strong elastocaloric effect, which can be used for next generation heating and cooling applications [7]. Notably, elastocaloric cooling is listed as one of the few disruptive technologies for managing the energy transition by the US Department of Energy [8]. The phase transition from the low-temperature martensite to the high-temperature austenite phase is well investigated in terms of its structural and morphological changes [9], including the formation of intermediate (R-) phases [10] or a strain glass phase [11]. However, these investigations mainly use polycrystalline samples [12]. It is advantageous to use single crystals in order to improve the accuracy of future measurements, to better investigate crystallographic effects, and to enable a more reliable comparison with model systems and theory. In the literature, single crystals of nickel-titanium are often grown using the Czochralski or Bridgman method. Typical growing speeds are in the range of several mm/h. Using these techniques, crystal dimensions of up to 30 mm in thickness and 50 mm in length have been achieved [13, 14]. To our best knowledge, no NiTi single crystals have been produced using the micro-pulling-down (µPD) method so far. This method also allows for easy variation of the alloy composition and doping because of the overall fast process. Here, our main goal is therefore to demonstrate the application of the µPD technique for the growth of NiTi single crystals. To create a flexible and easily adaptable system with full knowledge of all components, this µPD setup was self-built. The design freedom made it possible to rely largely on standard parts, resulting in a cost-effective system. The method is characterized by pulling the crystal downward through an opening at the bottom of a small crucible, which is heated by Joule or induction heating, onto a seed wire or seed crystal. Historically, µPD was developed to grow oxide optical fibers as single crystals [15]. However, numerous materials have since been produced using µPD. An overview of operated µPD systems is presented in Ref. [2]. Further setups were established for the growth of $LiMgPO_4$ [16], CuAlNi [17], $(Lu_{1/6}Y_{1/6}Ho_{1/6}Dy_{1/6}Tb_{1/6}Gd_{1/6})_3Al_5O_{12}$ [18, 19]. Here in contrast to more common methods typical pulling speed is in the range of 100 mm/h. Even though µPD is less common there are some available systems on the market. Some of the suppliers and their µPD setups are listed in Table 1.

**Table 1:** Micro-pulling-down setups available for purchase from companies

| Company | Heating | Speed | |
|---|---|---|---|
| Etudes et Constructions Mécaniques (ECM) | Resistive / inductive 2200 °C | 0.01 - 6000 mm/h | [20, 21] |
| Dai-ichi Kiden Co.,Ltd. (KDN) | Res. 1700 °C 6 kW | | [22] |
| | Hf induct. 2400 °C | 3 - 12000 mm/h | [23] |
| Crylink | IF 20 kW, 6-16 kHz | 0.1 - 6000 mm/h | [24] |

Here, we report on a custom-built µPD system granting access to all process parameters governing the growth process beyond the constrains of commercially available systems, including a step-by-step guide for its reconstruction and implementation.

## 2. Experimental

2.1 Scanning Electron Microscopy and Energy-Dispersive X-ray Spectroscopy

Scanning electron microscopy (SEM) and ion imaging were performed using a FEI Helios Nanolab 600 dual-beam electron microscope (FEI Germany, now Thermo Fisher Scientific), operated at 20 kV electron acceleration voltage for SEM imaging and 30 kV Ga⁺ ions for ion-beam imaging. Energy-dispersive X-ray spectroscopy (EDX) mapping was acquired at an electron acceleration voltage of 20 kV.

## 2.2 Electron Backscatter Diffraction

Electron backscatter diffraction (EBSD) measurements were performed with an FEI Quanta 650 FEG SEM equipped with a field emission gun operating at 30 kV. EBSD data were obtained using an EDAX-TSL system with a Hikari-XP camera, within a scanning area of 291,600 µm$^2$ and a step size of 1 µm. This leads to a total number of 292,681 detected Kikuchi patterns across the scanned region. Details on sample preparation for EBSD-analysis are documented in ref. [25].

## 2.3 Transmission Electron Microscopy

For the transmission electron microscopy (TEM) analysis, a JEM 2200 FS from JEOL was used, which was operated at 200 kV. The images were captured with a 4kx4k CMOS camera system, OneView from Gatan. The lamella was cut using a focused ion beam (FIB) from a random orientation and was thinned for TEM investigation.

## 2.4 Differential Scanning Calorimetry

To analyze the phase transition of the as grown NiTi Sample a Discovery DSC25 DSC from TA Instruments was used. The measurement was performed with 10 K/min in N$_2$ atmosphere. After the heat treatment, a NETZSCH DSC 204 F1 was used.

## 2.5 X-ray Photoelectron Spectroscopy

The NiTi sample was analyzed by X-ray photoelectron spectroscopy (XPS) using a *PHI VersaProbe* III device at 5×10$^{-10}$ mbar using monochromatic Al Kα irradiation and a hemispherical electron analyzer in constant analyzer energy mode. The angle between the sample plane and analyzer axis was 90° and the angle between source and analyzer was 45°. The spectra were recorded with a pass energy of 55 eV. The measured data was analyzed in CasaXPS Version 2.3.22. To compensate for charging effects and the unknown work function of the sample, the spectra were shifted so that the Fermi cutoff in the Ti/Ni 3d valance band coincides with a binding energy of 0 eV [26, 27]. Slight mismatches in binding energy positions between different spectra are likely due to differences in work functions. The main emission peaks of metallic Ni 2p were fitted with a Lorentzian asymmetric line shape $LA(\alpha,\beta,m)$, where $\alpha$ and $\beta$ are asymmetry parameters controlling the spread of the left and right peak tails, respectively, and the parameter $m$ defines the width of the convoluted Gaussian. The best fit was achieved by $LA(0.7,1.25,50)$ on a spline Tougaard background. The main emission peaks of metallic Ti 2p were fitted using a modified Lorentzian asymmetric line shape $LF(\alpha,\beta,w,m)$ with an additional tail damping factor $w$. The best fit was achieved by $LF(0.45,1.5,90,30,8)$ on a universal cross-section Tougaard background for the spectrum of the metallic sample (the same background extends to the O 1s peak, illustrating its accuracy). A Shirley background was used for Ti 2p of the oxidized sample. The carbide contribution to C 1s was fitted by $LA(1,4,150)$. All other component peaks are symmetric Voigt functions. The intensity ratios of all spin-orbit split peak doublets (including the satellites) were fixed to 1:2 for *p*-orbitals. Scofield cross sections were used for the quantification of the stoichiometry based on the narrow-scans [28].

2.6 Hot Gas Extraction Method

The oxygen content of the samples was analyzed via the hot gas extraction method (ONH-p2, ELTRA, Haan - Germany). The sample is heated in a graphite crucible to up to 3000 °C under inert gas, any oxygen present in the sample will react with the crucible to form CO or $CO_2$. The CO is converted to $CO_2$ in a catalysis oven, and the $CO_2$ content of the off-gas is measured via the extinction of an infrared laser at 4.3 µm wavelength. By the total $CO_2$ intensity measured over the 50 seconds of sample heating, and the provided sample mass, the oxygen content is determined via a calibration line derived from standard materials. The carbon content was measured by a very similar method, also based on the hot gas extraction (CS-i, ELTRA, Haan – Germany). Here the sample is heated inductively in an oxidizing atmosphere to above 2000 °C, and again the $CO_2$ formed by the combustion is measured by infrared absorption.

# 3. Results

## 3.1 Custom-Built µPD Setup

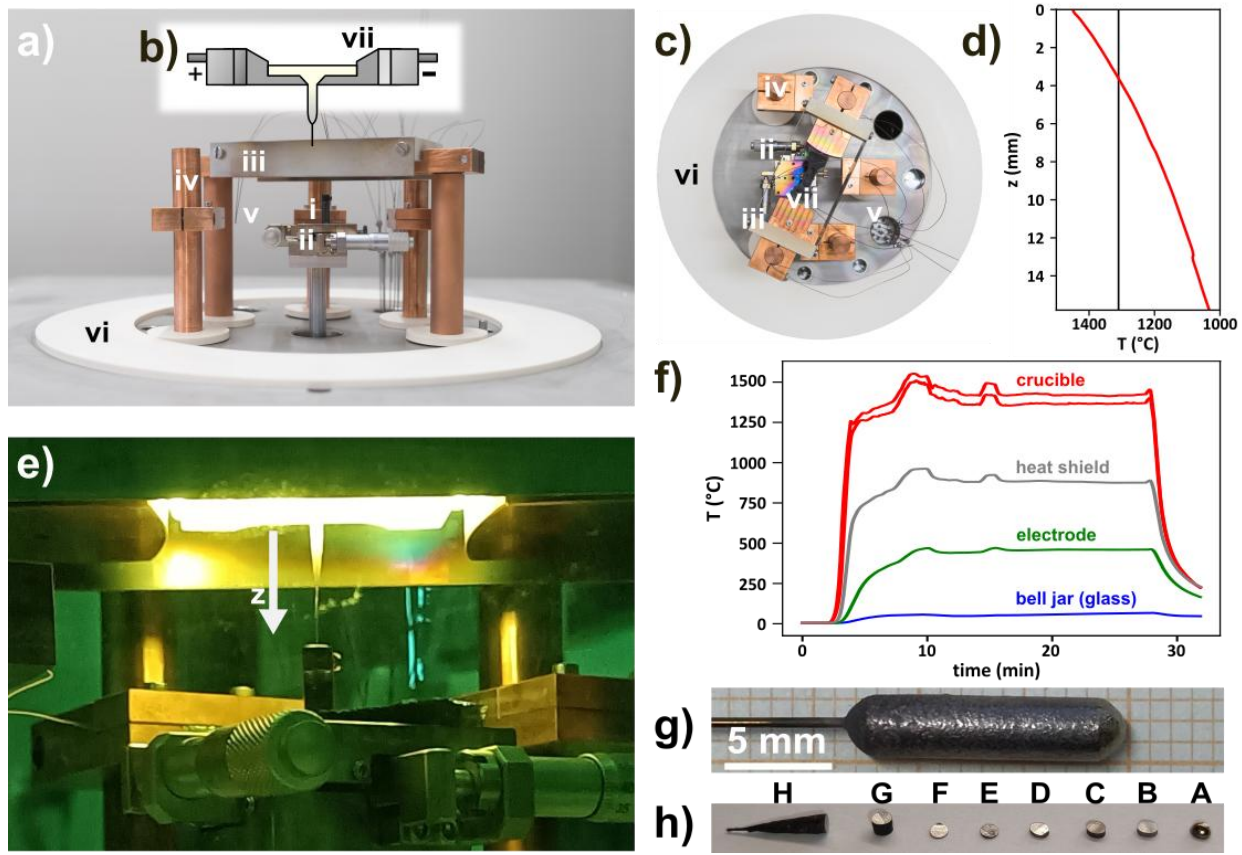

*Figure 1:* a) µPD setup with the process vacuum chamber, consisting of i. a wire holder, ii. a x-y-stage, iii. heat shielding, iv. a power feed threw, v. thermocouples, vi. a silicon gasket for a bell jar, and vii) the crucible; additional pictures can be found in Figure S1 b) schematic illustration of µPD; c) top view of the setup seen in a); d) temperature gradient measured along the pulling direction z during growth; e) photo taken during the pulling process; f) temperature at various positions during a typical growth process; g) photo of the sample used for TEM analysis; h) Investigated sample after cutting used for DSC, XPS, EBSD, SEM, EDX.

A µPD system, as shown in Figure 1a, consists of a crucible with a defined hole or capillary in the center (Figure 1c), a heating unit and a linear drive. In the case of µPD-grown oxides, which have been studied in the literature [29], platinum is typically used as crucible material. Due to the unavoidable disintegration of Pt in contact with NiTi melt, we use graphite crucibles. While graphite keeps its structural integrity when in contact with NiTi melt, it acts as a source of carbide impurities.

When using our graphite crucibles for NiTi, tests showed a capillary is not feasible, because a minimum diameter of 2mm is necessary to pull the melt. The geometry of the hole has a direct effect on whether or how well the melt can be drawn, due to the difference in surface tension and wetting properties of each material and even dopants. To pull the solidifying melt through the hole of the crucible, the melt must be attached to the linear drive, for this a seed wire can be used. The pulling speed and the corresponding cooling rate at a fixed crucible temperature influence the diameter of the crystal. The following table provides details of the components used for the custom-built µPD system.

**Table 2:** Components used for the self-made µPD assembly

| Component | Specifications | Company |
| --- | --- | --- |
| Linear drive | v = 0.06 - 2700 mm/h | HTM Reetz GmbH |
| DC power supply crucible | 1000 A @ 10 V | TDK-Lambda Germany GmbH |
| DC power supply secondary | 200 A @ 6 V | TDK-Lambda Germany GmbH |
| Turbopump station | 1,2 m$^3$/h | Edwards GmbH |
| Bell jar (glass) | hxb = 500 x 315 mm | DURAN Produktions GmbH & Co. KG |
| Thermocouples | Typ C | Omega Engineering GmbH |
| XY stage | l = ±6,5 mm | GMT Europe GMBH |
| Power cables | 1000 A | SCHRACK TECHNIK GMBH |
| Crucible | Graphite | Umicore Thin Film Products AG |
| Silicon gasket | 60 HT +280 °C | PLAZURA® René Höllrigl & J.Ahrends GbR |
| Vacuum parts and gaskets | FKM | Pfeiffer Vacuum Components & Solutions GmbH |
| Power feedthroughs | 1000 A 8 kV water cooled | tectra GmbH |

The chamber consists of a stainless-steel base plate with several welded flanges for the feedthroughs for the electrodes, linear drive, vacuum pump and temperature sensors (see Figure 1a,c). A glass bell jar is placed on a high-temperature-resistant silicone seal (Figure 1, vi), which is placed on the stainless-steel base plate. The glass bell jar is equipped with a gas supply. Technical drawings can be found in the supporting online material (Figure S2 - S4). The crucible is heated with a DC power supply, which can generate currents up to 1000 A at 10 V. The crucible is surrounded laterally by a stainless-steel heat shield (Figure 1, iii) for better temperature stabilization. This heat shield also helps to reduce the amount of heat radiation to which the outer glass bell jar is exposed. The holder for the heat shield is made of a temperature-resistant, machinable ceramic. We used an x-y-translation stage for positioning the seed crystal or seed wire under the hole in the crucible. In this case, it is important to ensure that the stage is greased with a vacuum grease. In addition, it must be positioned at a sufficient distance from the crucible to prevent thermal degradation of the fine adjustment springs of the x-y-translation stage. A smooth linear drive was achieved by a stepper-motor combined with a 17:1 reduction angular gearbox turning a spindle with a pitch of 5 mm along a linear track, resulting in a pulling speed between 0,06 mm/h and 2700 mm/h. The seed material for our experiments was titanium and nitinol wire (d=0.5 mm,

Ni$_{56.15}$Ti$_{43.77}$ from HME-Tech GmbH). The temperature was measured with type C thermocouples, as shown in Figure 1, v. The temperature was measured on the crucible, the electrodes, the heat shield and on the glass bell jar. It is essential to establish a stable temperature gradient along the pulling direction between the crucible and the crystal [2]. The temperature gradient, as seen in Figure 1d, was measured along the pulling direction starting at the crucible ($z = 0$ mm). The five power feedthroughs are arranged around the motor (Figure 1, iv), two of which are used for the crucible. The power feedthroughs are water cooled, the gas and waterflow diagram is presented in supplementary Figure S5. The 200 A power supply is connected to the other three feedthroughs, one directly and two via a switch. With the switch two can be connected individually or both simultaneously. A wiring diagram is provided in Figure S6. It is usually reported in the literature that the temperature gradient can be fine-tuned with an after-heater (a spring-like wire) to adjust the temperature decrease along the pulling direction [2]. We use these electrode feedthroughs for passing a DC current through a Ti wire, functioning as an oxygen getter instead of using it as an after-heater. The nominal melting temperature of NiTi is 1310 K [30], indicated by the vertical line of Figure 1d. To achieve temperatures of the melt well above this melting temperature, the graphite crucible was heated with a power of 1kW. Due to varying parameters, the temperatures of the melt in the crucible can differ greatly, measured by the thermocouple directly on the graphite crucible. Temperature measurements at the crucible, at an electrode, at the heat shield and at the glass bell jar are featured in Figure 1f. A uniform growth without significant thickness variation can be seen in Figure 1g. Due to direct resistive heating, only conductive crucible materials are suitable. Since metallic crucibles react with NiTi melts, graphite is the only viable option. However, in the case of graphite, small amounts of carbon are unavoidably dissolved in the NiTi melt. The carbon pickup of liquid NiTi is low when temperatures and process / exposure durations are kept small [31]. In fact, using graphite crucibles represents a standard approach in research environments and also in shape memory industry when NiTi alloys are prepared [32]. Already in the 1960s, Drennen et al. [33] suggested graphite as a crucible material as other potential candidates shows significantly intensive reactions. However, carbon pickup should be kept low, because it promotes the formation of carbides of type TiC, which affects transformation temperatures as due to changes in the Ni/Ti ratio[34, 35]. In general, NiTi SMAs react sensitive to pick of small amounts of impurities like carbon and oxygen, which are typically contained in raw materials, or which are introduced into the material during processing, e.g. [34, 36, 37]. The crucible that we used has a filling volume of approx. 0.5 cm$^3$. Holes were drilled in the center of the crucible and were also countersunk to mitigate the 2 mm thickness of the crucible bottom. The crystal growth was executed in a vacuum after purging with argon, and using a heated titanium wire as an additional oxygen getter. The Ti wire must be changed after each experiment to ensure a sufficient O$_2$ uptake, otherwise the wire will get brittle over time. Operational pressures of $10^{-8}$ mbar were achieved. With regard to the total cost of the system, one must assume that the costs of the components vary greatly and are also continuously increasing. But in our case, the chamber could be constructed in the university's own workshop. Therefore, the power source, the linear drive and the turbo molecular vacuum pump proved to be the most expensive components of the setup.

## 3.2 Sample Preparation

NiTi fragments of polycrystalline $Ni_{50}Ti_{50}$ were used as the starting material, which had been pre-synthesized by an optimized arc melting procedure documented in Ref. [34]. The starting materials, Ni-pellets and Ti-blocks used both had a purity of 99.995 wt.% (metal basis); both were obtained from Hauner Metallische Werkstoffe, Röttenbach, Germany. The crucible was filled with fragments totaling a weight close to 10 g. A first sample grown with a pulling speed of 327 µm/min just below 1600 °C using a Ti wire was cut into discs perpendicular to the pulling direction for further investigation, as shown in Figure 1h. For the cuts a diamond wire saw was used and polished with up to 0,25 µm polishing suspension. The cut **G** was used for DSC and XPS and SEM with EDX. The cut **C** was used for EBSD with additional electro polishing.

## 3.3 Morphology and Microstructures

The microstructure of the as-grown crystal was investigated by SEM. Figure 2a presents an overview back scatter micrograph where three different constituents can be identified, based on contrast and results documented in previous studies [32, 34, 35, 38], where detailed TEM and SEM analysis were conducted. The grey matrix represents binary NiTi which can be either austenite or martensite, depending on temperature. The slightly darker phase represents $Ti_2Ni$ with small oxygen in solid solution [35, 39]. The third dark grey phase in Figure 2a represents a large carbide of type TiC, as will be discussed. EDX mappings of Ni and Ti are presented in Figure 2b. The Ti level of the matrix was detected as 52 at.% and the slightly darker phase contained close to 64 at.% Ti, which both confirms the conclusions made above. For the third phase, only Ti and no Ni was detected. A close look at the low-energy part of the EDX spectrum reveals the presence of carbon. We assume that this phase is a carbide phase, in line with thermodynamic data published in literature [40].

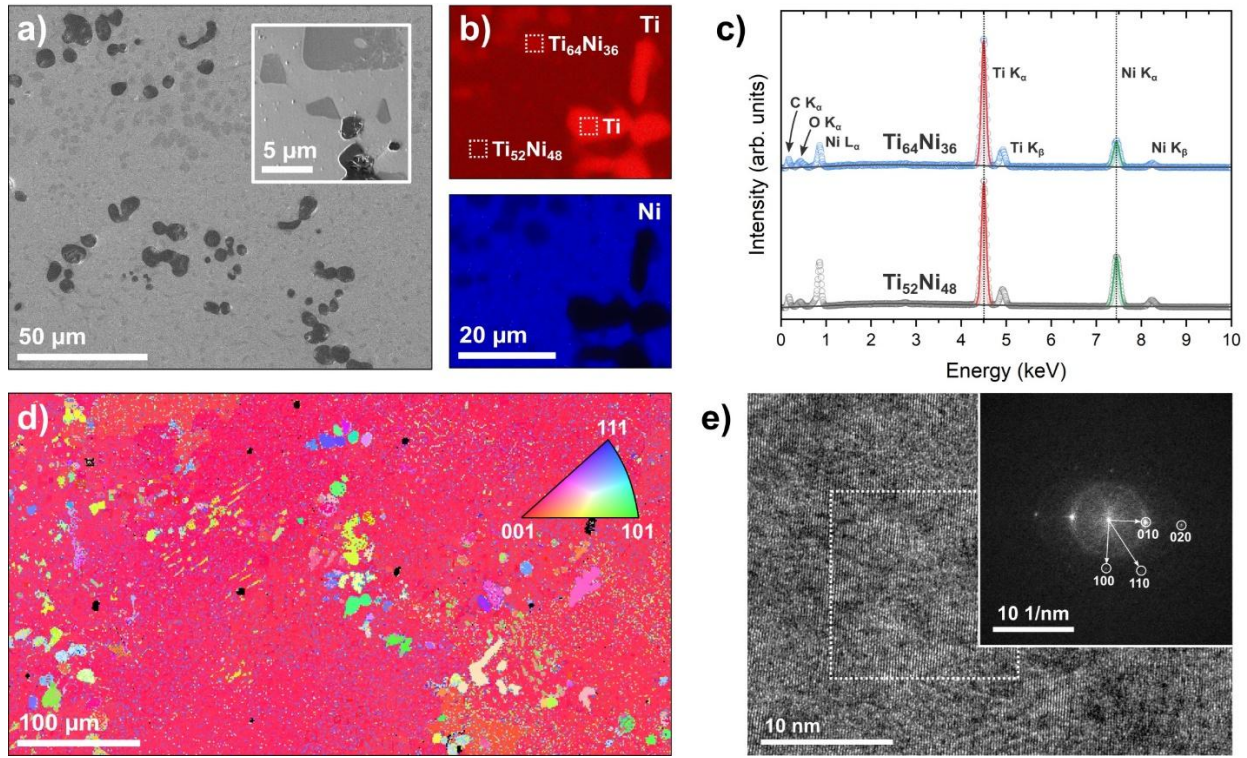

***Figure 2:*** *a) SEM images showing inclusions on the surface, with the inset taken by Ga⁺ ion beam imaging b,c) EDX mapping and point analysis on the marked areas show that the NiTi composition is not completely homogeneous over the whole surface d) EBSD mapping indicating a (001) orientation e) TEM bright field image of the NiTi lamella with FFT inset.*

We used EBSD to assess crystalline quality of the material Figure 2 d, with the corresponding inverse pole figure (IPF) orientation mapping. The dominating red color shows that most parts of the material are characterized by an (001) orientation, i.e. the 001 plane normal is parallel to the longitudinal axis of the as-grown crystal. However, the microstructural region shown in Figure 2d does not have a uniform orientation spread. In fact, some regions are characterized by orientations which significantly differ from those of the NiTi matrix. These regions correspond to the carbides of type TiC, which can be present in NiTi melts [31] (dark grey phase in Figure 2a). These particles might have formed as primary carbides in the early stages of solidification, or, they might have formed at the interfaces between graphite crucibles and NiTi melts, from where they can float into the liquid alloy [31]. We note that the appearance of impurity-related phases depends on duration in which the molten alloy is exposed to graphite and atmospheres with residual oxygen and therefore pulling speed. Nevertheless, the appearance of these constituent is not unusual. Even highest purity NiTi SMAs contain both $Ti_2Ni$ (with O) and TiC, as oxygen and carbon cannot be solved in NiTi [36]. In a next step TEM was used to characterize a second but comparable sample (nitinol wire as seed with v = 60 mm/h at approx. 1366 °C). With high resolution TEM images, long range crystallinity of the NiTi sample has been observed. Small, intermediate areas show an amorphous structure, which might be caused by the preparation and thinning of the lamella. The inset shows the Fast Fourier Transform (FFT) of the region indicated by the dotted line. The diffraction spots are assigned to the monoclinic martensite phase.

## 3.4 Chemical Composition

It is common practice to estimate the Ni/Ti ratio from the peak temperatures observed in DSC measurements. This connection was found empirically and is described in detail in Ref. [35]. From the DSC data, we obtain an overall composition of $Ni_{50.8}Ti_{49.2}$ using the martensite start temperature $M_S$. To verify the DSC measurement the sample was heat treated. This was done at 900°C for 5 h with a constant flow of argon and later quenched in water. The comparison of the two DSC measurements is shown in Figure 3a. In the as-grown sample precipitations are present, where after heat treatment these are dissolved and therefore give a slightly different peak shape but with only a -1 K $M_S$ change.

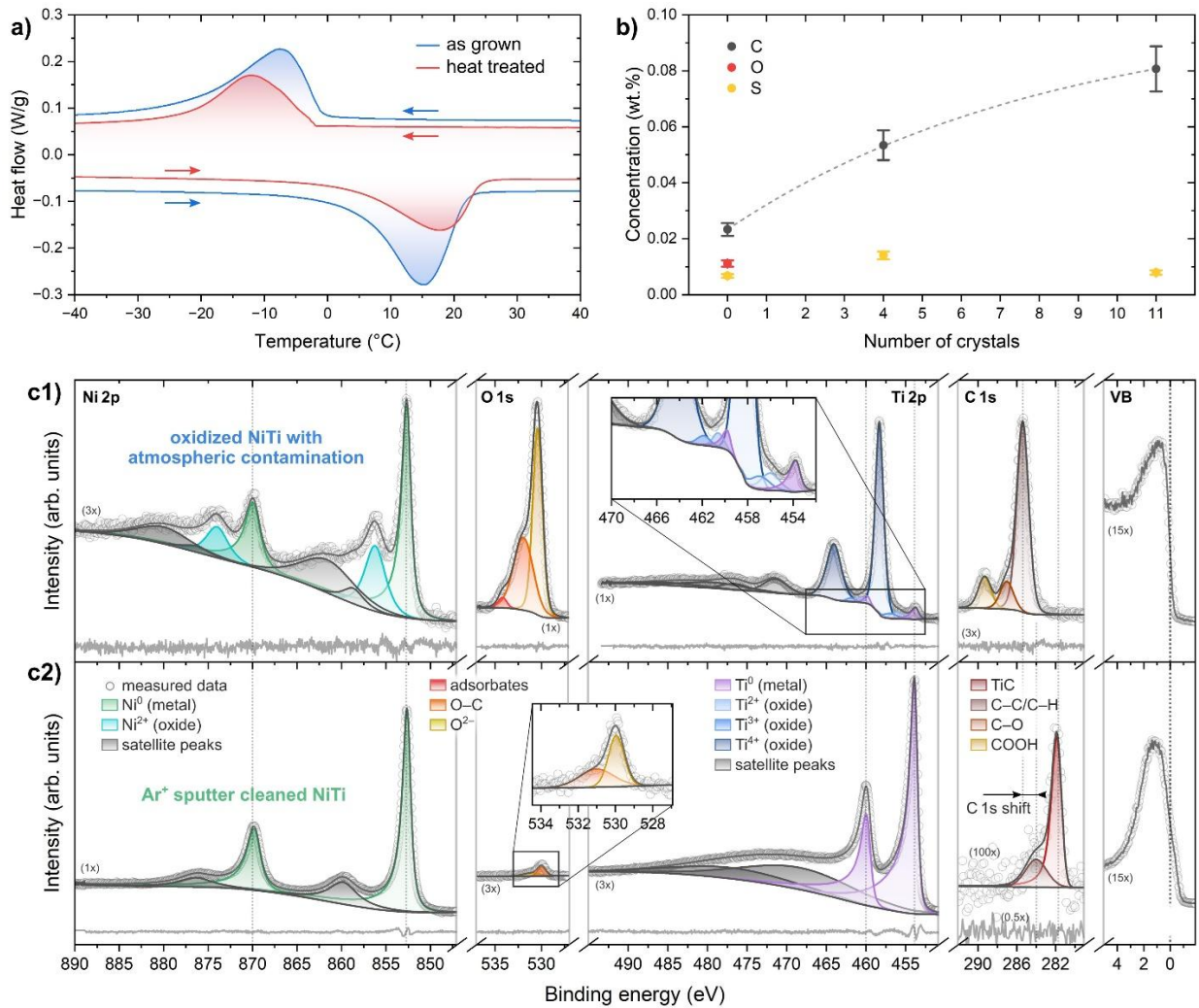

*Figure 3:* a) Heat flow of the sample at 10 K/min b) Increase of carbon (wt.%) in NiTi Samples prepared from a refilled melt analyzed by the hot gas extraction method. c1) XPS measurements of the NiTi sample before and c2) after sputter etching. Some peak regions have been enlarged for better readability by the factors shown in the diagram right below the data. The fit residuals are shown in pale grey below the spectra.

We therefore used XPS to characterize the quantities of carbon and oxygen incorporated into the sample, as well as the characteristic nickel and titanium bond types. The native oxide layer, as characterized by the top XPS spectrum in Figure 3c, is dominated by Ti due to its higher oxygen affinity compared to Ni. The nominal composition resulting from the shown peak fitting models is as follows: 5.0 at.% Ni, 15.3 at.% Ti, 38.0 at.% C, and 41.8 at.% O. The Ti signal mostly originates from Ti oxides, namely $Ti^{2+}$, $Ti^{3+}$, and $Ti^{4+}$ ($TiO_2$), where the latter is most prevalent, which is typical for native Ti oxide layers [41]. Ni is only partially oxidized at the surface. The composition accounted for adventitious carbon is as follows: 9.7 at% Ni (6.6 at.% $Ni^0$ and 3.1 at.% $Ni^{2+}$), 36.4 at.% Ti (2.3 at.% $Ti^0$, 31.3 at.% $Ti^{4+}$, 1.5 at.% $Ti^{3+}$, and 1.3 at.% $Ti^{2+}$), and 53.9 at.% $O^{2-}$. Note that the uncertainties of these values strongly depend on the ambiguity of the fitting models, which is hard to quantify, but likely higher than $\pm 5\%$ (not at.%) of the stated values. The oxidized surface was etched away by $Ar^+$ sputtering to investigate the bulk properties. The Ni and Ti signals are apparently fully metallic. Most of the carbon contamination is in the form of carbides (likely TiC) with very little contribution of oxygenated hydrocarbons. The composition is as follows: 40.9 at.% Ni, 50.5 at.% Ti, 6.4 at.% C (4.0 at.% TiC and 1.4 at.% C-$sp^3$), and 3.0 at.% O (1.3 at.% $O^{2-}$ and 1.7 at.% O–C). Mind that the quantitative compositions are averaged over the information depth and can be influenced by artifacts like selective sputtering. The deviation in Ni to Ti concentration can be traced back to the Ti and Ni rich areas, seen by EDX and EBSD in Figure 2. The deficit in titanium in the sample seems to arise from the formation of titanium-carbides and oxides during growth which are randomly distributed throughout the sample and are therefore detected differently by DSC and XPS. The formation of carbides from the melt-crucible interface and preexisting carbon in the melt material also affects the Ni:Ti ratio [34]. Therefore, we have characterized the carbon absorption by the melt and the subsequent carbon uptake into the samples. To do this, we repeatedly melted the same melt and refilled it, so that the carbon content of the melt increased from experiment to experiment. As an example, we examined the carbon concentration for the 4[th] and 11[th] sample. Carbon and oxygen content of the samples were analyzed via the hot gas extraction method. Figure 3b shows the resulting quantitative proportion of the carbon and oxygen content in the samples. There is a clear correlation between carbon uptake and dwell time. It can be concluded that the residence time of the melt in the graphite crucible must be as short as possible to minimize the carbon uptake. Re-melting for follow-up experiments cannot be recommended if carbon contamination is to be avoided. Instead, it is necessary to use a new crucible for each experiment to keep the carbon uptake as small as possible.

## 4. Conclusion

In this work, we designed, built, and tested a custom µPD system for the growth of NiTi single crystals. Our results establish µPD as a fast, flexible, and cost-efficient method for fabricating oriented NiTi single crystals, enabling systematic studies of phase transitions and dopant effects. All major components of the apparatus are listed in Section 2. Controlling the gas atmosphere is particularly important for the production of NiTi, as NiTi is extremely oxygen-affine when molten. Even under high vacuum ($10^{-8}$ mbar) and using an oxygen getter, approximately 3% oxygen was still detected in the NiTi sample after the growth process. We observed a significant carbon uptake, which manifested itself as titanium carbide precipitation in the crystal. Therefore, the contact time between the melt and the graphite crucible is a critical factor for the quality of the grown crystals and should be minimized in order to obtain high-quality NiTi crystals. EBSD measurements show that the µPD grown crystals are predominantly (001) oriented parallel to the growth direction. The developed setup expands the accessible parameter space for NiTi growth and opens new opportunities for tailoring material properties, ultimately paving the way for advanced investigations of phase-transition mechanisms and functional optimization in shape memory alloys.

# Declaration of Generative AI and AI-assisted technologies in the writing process

During the preparation of this work the authors used AI tools in order to improve the readability and language of the manuscript. After the use of AI, the authors reviewed and edited the content as needed and take full responsibility for the content of the published article.

# CRediT authorship contribution statement

**T. Sieweke:** Investigation, Writing. **C. Luther:** Investigation. **M. Wortmann:** Investigation, Writing. **L. Schnatmann:** Supervision. **F. Werner:** Investigation. **O. Kuschel:** Investigation. **L. Bondzio:** Investigation. **I. Ennen:** Investigation. **J. Bünte:** Investigation. **K. Rott:** Investigation. **O. Oluwabi:** Investigation. **M. Loewenich:** Investigation. **J. Wollschläger:** Supervision, Resources. **A. Hütten:** Supervision, Resources. **J. Frenzel:** Resources, Writing. **G. Schierning:** Conceptualization, Resources, Writing. **A. Kunzmann:** Conceptualization, Supervision, Writing.

# Data availability

The data that support the findings of this study are available from the corresponding author upon reasonable request.

# References

[1] A. Grünebohm, A. Hütten, A.E. Böhmer, J. Frenzel, I. Eremin, R. Drautz, I. Ennen, L. Caron, T. Kuschel, F. Lechermann, D. Anselmetti, T. Dahm, F. Weber, K. Rossnagel, G. Schierning, A Unifying Perspective of Common Motifs That Occur across Disparate Classes of Materials Harboring Displacive Phase Transitions, Adv Energy Mater, 13 (2023).
[2] T. Fukuda, V.I. Chani, Shaped crystals: growth by micro-pulling-down technique, Springer Science & Business Media, 2007.
[3] F.E. Wang, W.J. Buehler, S.J. Pickart, Crystal Structure and a Unique Martensitic Transition of TiNi, J Appl Phys, 36 (1965) 3232-+.
[4] D. Stoeckel, A. Pelton, T. Duerig, Self-expanding nitinol stents: material and design considerations, Eur Radiol, 14 (2004) 292-301.
[5] H. Yuan, J.C. Fauroux, F. Chapelle, X. Balandraud, A review of rotary actuators based on shape memory alloys, J Intel Mat Syst Str, 28 (2017) 1863-1885.
[6] D.J. Hartl, D.C. Lagoudas, Aerospace applications of shape memory alloys, P I Mech Eng G-J Aer, 221 (2007) 535-552.
[7] S.M. Kirsch, F. Welsch, N. Michaelis, M. Schmidt, A. Wieczorek, J. Frenzel, G. Eggeler, A. Schütze, S. Seelecke, NiTi-Based Elastocaloric Cooling on the Macroscale: From Basic Concepts to Realization, Energy Technol-Ger, 6 (2018) 1567-1587.
[8] H. Mevada, B.Y. Liu, L. Gao, Y. Hwang, I. Takeuchi, R. Radermacher, Elastocaloric cooling: A pathway towards future cooling technology, Int J Refrig, 162 (2024) 86-98.

[9] A.L. Roitburd, G.V. Kurdjumov, Nature of Martensitic Transformations, Mater Sci Eng, 39 (1979) 141-167.
[10] X.B. Wang, B. Verlinden, J. Van Humbeeck, R-phase transformation in NiTi alloys, Mater Sci Tech-Lond, 30 (2014) 1517-1529.
[11] Q.L. Liang, D. Wang, X.D. Ding, Y.Z. Wang, Two-step strain glass transition in NiTi shape memory alloy with unique properties, Mater Res Lett, 12 (2024) 678-687.
[12] A. Kunzmann, J. Frenzel, U. Wolff, J. Han, L. Giebeler, D. Piorunek, M. Mittendorff, J. Scheiter, H. Reith, N. Perez, K. Nielsch, G. Eggeler, G. Schierning, The role of electrons during the martensitic phase transformation in NiTi-based shape memory alloys, Mater Today Phys, 24 (2022).
[13] T.M. Brill, S. Mittelbach, W. Assmus, M. Mullner, B. Luthi, Elastic Properties of NiTi, J Phys-Condens Mat, 3 (1991) 9621-9627.
[14] T. Simon, A. Kröger, C. Somsen, A. Dlouhy, G. Eggeler, In-situ TEM cooling/heating experiments on deformed NiTi shape memory single crystals, in: European Symposium on Martensitic Transformations, EDP Sciences, 2009, pp. 02030.
[15] D.H. Yoon, I. Yonenaga, T. Fukuda, N. Ohnishi, Crystal growth of dislocation free $LiNbO_3$ single crystals by micro pulling down method, J Cryst Growth, 142 (1994) 339-343.
[16] D. Kulig, W. Gieszczyk, P. Bilski, B. Marczewska, M. Klosowski, New OSL detectors based on $LiMgPO_4$ crystals grown by micro pulling down method. Dosimetric properties vs. growth parameters, Radiat Meas, 90 (2016) 303-307.
[17] I. López-Ferreño, J.S. Juan, T. Breczewski, G.A. López, M.L. Nó, Micro pulling down growth of very thin shape memory alloys single crystals, Funct Mater Lett, 10 (2017).
[18] M. Pianassola, M. Loveday, B.C. Chakoumakos, M. Koschan, C.L. Melcher, M. Zhuravleva, Crystal Growth and Elemental Homogeneity of the Multicomponent Rare-Earth Garnet $(Lu_{1/6}Y_{1/6}Ho_{1/6}Dy_{1/6}Tb_{1/6}Gd_{1/6})_3Al_5O_{12}$, Cryst Growth Des, 20 (2020) 6769-6776.
[19] M. Pianassola, K.L. Anderson, J. Safin, C. Agca, J.W. McMurray, B.C. Chakoumakos, J.C. Neuefeind, C.L. Melcher, M. Zhuravleva, Tuning the melting point and phase stability of rare-earth oxides to facilitate their crystal growth from the melt, J Adv Ceram, 11 (2022) 1479-1490.
[20] ECM, micro pulling down (mpd) crystal growth furnace, in, 07.02.2025, pp. https://ecm-greentech.fr/portfolio/micro-pulling-down-mpd-micro-pulling-down-crystal-growth-furnace/.
[21] ECM_LAB, Micro pulling down (MPD) : crystal growth furnace, in, 07.02.2025, pp. https://ecmlabsolutions.com/products/crystal-growth-mpd/.
[22] KDN, Resistance heating type μ-PD furnace, in, 07.02.2025, pp. https://www.d-kdn.co.jp/eng/product/c/033.html.
[23] KDN, High-frequency heating μ-PD furnace, in, 07.02.2025, pp. https://www.d-kdn.co.jp/eng/product/c/030.html.
[24] Crylink, CL-XDF Type Micro Pulling Down Single Crystal Furnace, in, 07.02.2025, pp. https://www.crylink.com/product/cl-xdf-type-micro-pulling-down-single-crystal-furnace/.
[25] P. Thome, S. Medghalchi, J. Frenzel, J. Schreuer, G. Eggeler, Ni-Base Superalloy Single Crystal (SX) Mosaicity Characterized by the Rotation Vector Base Line Electron Back Scatter Diffraction (RVB-EBSD) Method, Ultramicroscopy, 206 (2019) 112817.
[26] G. Greczynski, L. Hultman, Binding energy referencing in X-ray photoelectron spectroscopy, Nat Rev Mater, 10 (2025) 62-78.
[27] G. Greczynski, Binding energy referencing in X-ray photoelectron spectroscopy: Expanded data set confirms that adventitious carbon aligns to the sample vacuum level, Appl Surf Sci, 670 (2024).
[28] J.H. Scofield, Hartree-Slater subshell photoionization cross-sections at 1254 and 1487eV, J Electron Spectrosc, 8 (1976) 129-137.
[29] V.I. Chani, K. Shimamura, T. Fukuda, Flux growth of $KNbO_3$ crystals by pulling-down method, Cryst Res Technol, 34 (1999) 519-525.

[30] J.L. Murray, Phase Diagram Ni-Ti, in: T.B. Massalski, H. Okamato, P.R. Subramanian, L. Kacprzak (Eds.) Binary Alloy Phase Diagrams, ASM International, Materials Park, Ohio, 1996.
[31] Z.H. Zhang, J. Frenzel, K. Neuking, G. Eggeler, On the reaction between NiTi melts and crucible graphite during vacuum induction melting of NiTi shape memory alloys, Acta Materialia, 53 (2005) 3971-3985.
[32] J. Frenzel, Z. Zhang, K. Neuking, G. Eggeler, High quality vacuum induction melting of small quantities of NiTi shape memory alloys in graphite crucibles, Journal of Alloys and Compounds, 385 (2004) 214-223.
[33] D.C. Drennen, C.M. Jackson, H.J. Wagner, The development of melting and casting procedures for nitinol nickel-base alloys, Battelle Memorial Institute, 1968.
[34] J. Frenzel, Z. Zhang, C. Somsen, K. Neuking, G. Eggeler, Influence of carbon on martensitic phase transformations in NiTi shape memory alloys, Acta Materialia, 55 (2007) 1331-1341.
[35] J. Frenzel, E.P. George, A. Dlouhy, C. Somsen, M.F.X. Wagner, G. Eggeler, Influence of Ni on martensitic phase transformations in NiTi shape memory alloys, Acta Mater, 58 (2010) 3444-3458.
[36] M. Rahim, J. Frenzel, M. Frotscher, J. Pfetzing-Micklich, R. Steegmüller, M. Wohlschlögel, H. Mughrabi, G. Eggeler, Impurity levels and fatigue lives of pseudoelastic NiTi shape memory alloys, Acta Materialia, 61 (2013) 3667-3686.
[37] C. Haberland, M. Elahinia, J.M. Walker, H. Meier, J. Frenzel, On the development of high quality NiTi shape memory and pseudoelastic parts by additive manufacturing, Smart Materials and Structures, 23 (2014) 104002.
[38] J. Mentz, J. Frenzel, M.F.X. Wagner, K. Neuking, G. Eggeler, H.P. Buchkremer, D. Stover, Powder metallurgical processing of NiTi shape memory alloys with elevated transformation temperatures, Materials Science and Engineering A, 491 (2008) 270-278.
[39] M.V. Nevitt, Stabilization of Certain $Ti_2Ni$-Type Phases by Oxygen, Transactions of the American Institute of Mining and Metallurgical Engineers, 218 (1960) 327-331.
[40] Y. Du, J.C. Schuster, Experimental investigation and thermodynamic modeling of the Ni-Ti-C system, Zeitschrift für Metallkunde, 89 (1998) 399-410.
[41] M. Wortmann, K. Viertel, M. Westphal, D. Graulich, Y. Yang, M. Gaerner, J. Schmalhorst, N. Frese, T. Kuschel, Sub-Nanometer Depth Profiling of Native Metal Oxide Layers Within Single Fixed-Angle X-Ray Photoelectron Spectra, Small Methods, 8 (2024).

# Supplementary Material

Pictures of the Setup:

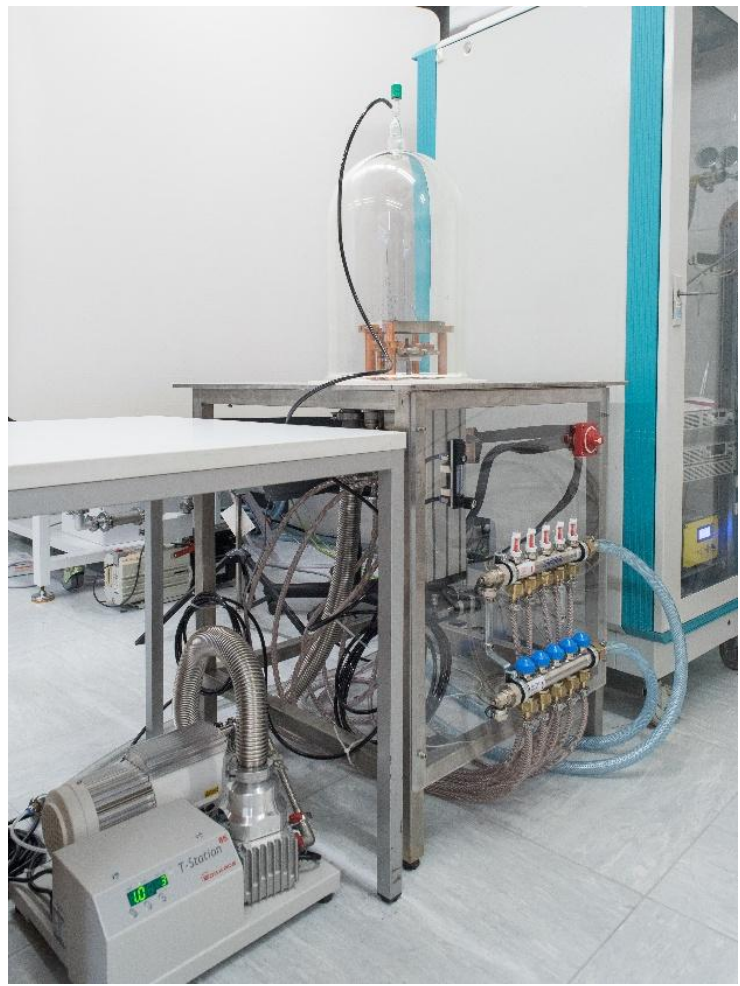

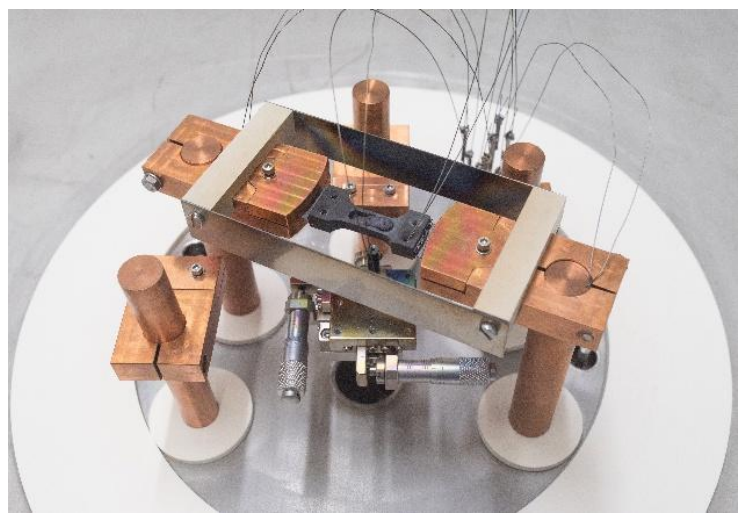

**Figure S1:** View of the whole system and closeup on the growing chamber.

Technical drawings:

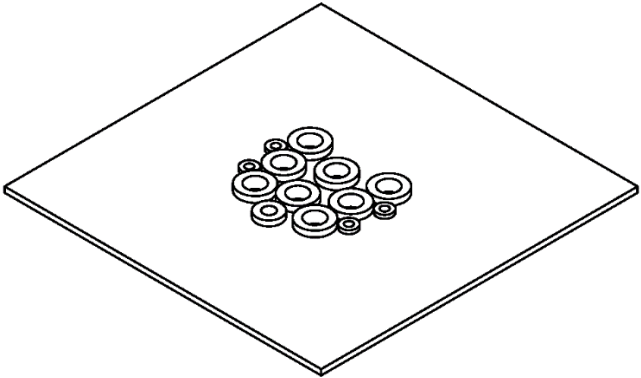

**Figure S2:** Baseplate with welded CF-Flanges, view from below.

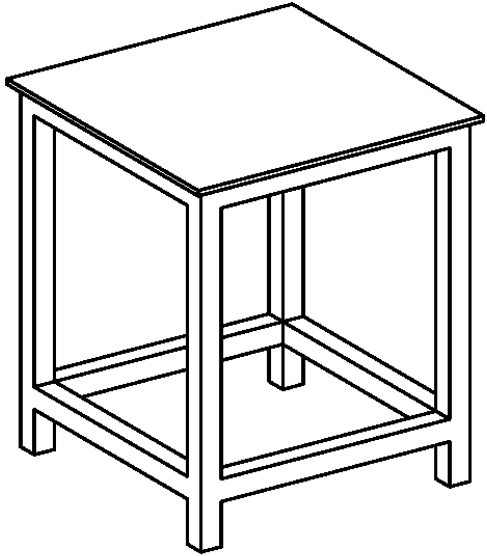

**Figure S3:** Table for Baseplate

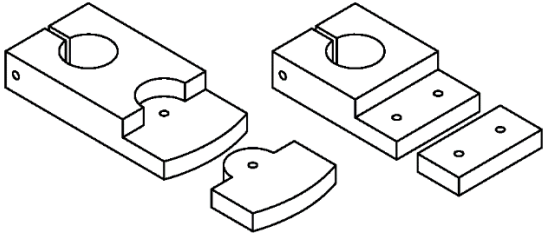

**Figure S4:** Connection brackets from power feedthrough to crucible

Flow diagram and electrical diagram:

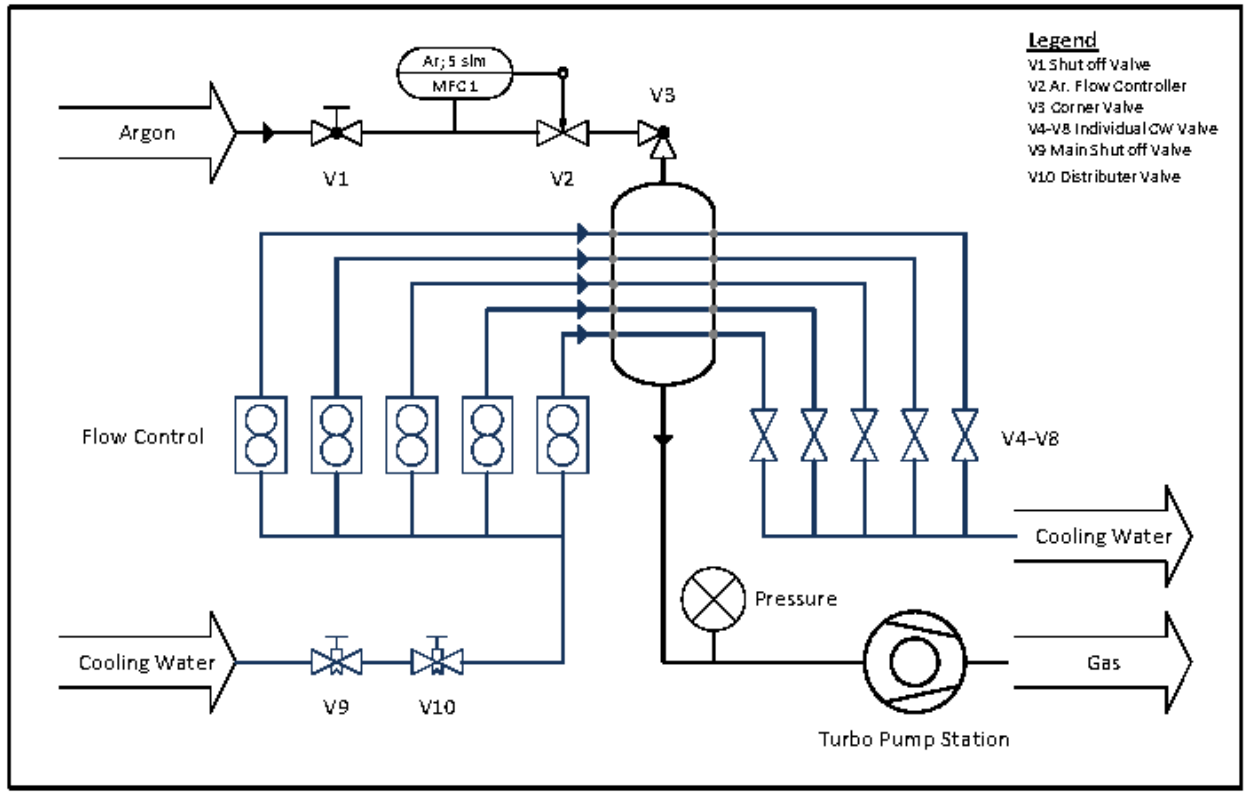

**Figure S5:** The gas and water flow diagram is added for the deeper understanding of the control and operation of the µPD system and the arising safety precautions.

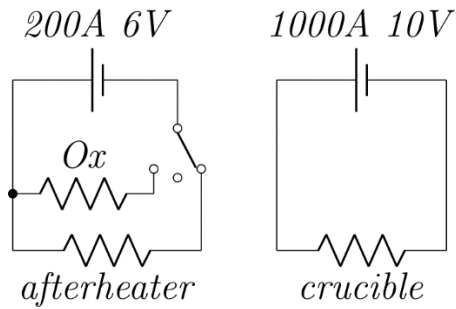

**Figure S6:** The wiring diagram is shown for both power supplies.